# An Improved Finite Element Modeling Method for Triply Periodic Minimal Surface Structures Based on Element Size and Minimum Jacobian

**Siqi WANG [a], Chuangyu JIANG [a], Xiaodong ZHANG [a], Yilong ZHANG [a], Baoqiang ZHANG [a,b]\*, Huageng LUO [a]\***

*aSchool of Aerospace Engineering, Xiamen University, Xiamen 361102, China*

*bAero Engine Academy of China, Beijing 101300, China*

**Abstract:** Triply periodic minimal surface (TPMS) structures, a type of lattice structure, have garnered significant attention due to their lightweight nature, controllability, and excellent mechanical properties. Voxel-based modeling is a widely used method for investigating the mechanical behavior of such lattice structures through finite element simulations. This study proposes a two-parameter voxel method that incorporates joint control of element size and minimum Jacobian (MJ). Numerical results indicate that the simulation outcomes tend to stabilize when the MJ reaches 0.3. The grid convergence index (GCI), based on Richardson extrapolation, is introduced to systematically assess the numerical convergence behavior of both voxel models and the proposed two-parameter voxel models. This provides a systematic and objective framework for evaluating discretization errors and mesh convergence in TPMS modeling. Compared with traditional voxel method, the proposed method exhibits superior mesh convergence, solution accuracy, and computational efficiency. Furthermore, the two-parameter voxel method also shows excellent applicability in the analysis of graded TPMS structures, exhibiting even better convergence behavior than in uniform structures.

## 1. Introduction

Lattice structures represent a class of periodic porous materials. Based on their geometric configurations, three-dimensional lattice structures can be categorized into three types: truss-based 3D lattice structures composed of strut elements, plate-based 3D lattice structures formed by flat panels, and shell-based 3D lattice structures constructed from curved surfaces [1]. Among them, triply periodic minimal surface (TPMS) structures are a typical representative of shell-based 3D lattice structures. TPMS structures are nature-inspired architectures [2] characterized by zero mean surface curvature [3], which distinguishes them from conventional lattice structures. This geometric feature enables a more uniform stress distribution under loading, resulting in superior load-bearing capacity [4]. Moreover, TPMS structures exhibit enhanced performance in impact resistance and energy absorption compared to other types of lattice structures [5], highlighting their broad potential for engineering applications.

Numerical simulation serves as a critical tool for investigating the mechanical behavior of lattice structures. Truss-based 3D lattice structures are typically discretized using beam elements [6] or hexahedral solid elements [7]. Plate-based lattice structures are commonly modeled with tetrahedral or hexahedral elements [8], both achieving reliable simulation accuracy. However, unlike these lattice types, TPMS structures exhibit unique curved surface geometries, making the construction of finite element models particularly crucial for ensuring both computational accuracy and efficiency. Currently, several finite element modeling methods are employed for TPMS structures, including shell models [9-11], tetrahedral models [12-13], hexahedral models [14], homogenized models [15], and voxel models [16-22]. Among these methods, generating high-quality tetrahedral or hexahedral meshes often requires extensive manual operations and the replication of meshed unit cells, which heavily rely on expert knowledge and experience [2]. For graded TPMS structures, the modeling process becomes even more labor-intensive, as each unit cell must be modeled individually [23]. Numerical homogenization is an essential tool for analyzing multiscale physical phenomena and is suitable for structures composed of a large number of TPMS cells. Nevertheless, it is highly dependent on finite element analysis, which imposes considerable computational costs [24]. The voxel modeling method discretizes TPMS structures using regular hexahedral solid elements. Due to its simplicity in implementation and broad applicability, voxel modeling has been widely adopted in the mechanical performance analysis of TPMS structures.

Several researchers have conducted comparative studies on existing finite element modeling approaches for TPMS structures. Qiu et al. [2] evaluated five different modeling methods—triangular shell models, quadrilateral shell models, tetrahedral solid models, hexahedral solid models, and voxel models—in terms of modeling complexity, computational efficiency, data management, and simulation accuracy. Their findings indicate that quadrilateral element shell models are better suited for TPMS structures with low relative density or thin walls, whereas voxel-based models are more appropriate for structures with higher relative density or thicker walls. Similarly, Simsek et al. [25] developed finite element models of TPMS structures using shell elements, solid elements, homogenization methods, super-elements, and voxel elements, and compared the modal characteristics obtained from each modeling approach.

The results showed that shell element models exhibited limited applicability, while solid element models involved the most labor-intensive meshing process and demonstrated the lowest computational efficiency. In contrast, voxel models offered the simplest modeling process, higher computational efficiency than solid models, and broad applicability across various TPMS configurations.

According to established standards for verification and validation, the first step in the verification process is a mesh refinement study [26], which aims to evaluate discretization errors and ensure that the developed mesh is sufficiently refined. In the context of mesh convergence analysis for voxel TPMS models, Maskery et al. [16] investigated the effects of mesh size, lattice type, orientation, and volume fraction on the effective Young's modulus. The study indicated that the Young's modulus can be considered converged when each unit cell contains approximately 50,000 voxel elements. In another study, Maskery et al. [17] defined mesh convergence as a variation in Young's modulus of less than 1%. The results showed that for 4×4×4 Gyroid and Diamond structures, approximately 3,000,000 voxel elements are required to reach convergence (about 47,000 elements per unit cell), while the Primitive structure requires around 2,000,000 voxel elements (about 31,000 elements per unit cell). Montazerian et al. [18] further established a quantitative relationship between the number of elements per unit cell and the error in the converged solution. Based on this relationship, they predicted that approximately 64,000 voxel elements per unit cell are required to achieve a 1% variation in the relative elastic modulus. In addition, the study emphasized that when evaluating the convergence of graded TPMS structures, the convergence criterion should be satisfied in the region with the lowest local relative density. Aremu et al. [19] conducted static analyses of Gyroid, body centered cubic (BCC), and face centered cubic (FCC) structures using simplified finite element models based on the voxel method. The study investigated the stiffness and mesh convergence behavior of these three lattice structures. For the Gyroid structure, they reported that convergence was achieved when each unit cell comprised approximately 250,000 elements. Wan et al. [20] investigated the energy dissipation capabilities of Primitive, Gyroid, Diamond, and I-Wrapped Package lattices under cyclic loading through both numerical simulations and experiments. In their study, each unit cell was discretized using approximately 8,100 voxel elements. However, in the numerical simulation models of TPMS structures developed by Afshar et al. [21] and Kadkhodapour et al. [22], each unit cell was discretized using approximately 4,600 and 4,000 elements, respectively. Another study indicated that for TPMS structures fabricated via additive manufacturing, the size of voxel elements can be determined based on the desired surface finish [27].

The above analysis indicates that there is considerable variation in the number of elements used in voxel models across different studies. Currently, there is no broad consensus regarding the appropriate mesh resolution for achieving convergence in voxel TPMS models. Analytical solutions of mathematical models are helpful in quantifying discretization errors. However, for TPMS structures, such analytical solutions are often difficult to obtain. In addition, conventional voxel-based TPMS modeling methods typically adopt a strategy in which the Jacobian of each element is equal to 1, meaning that TPMS structures are discretized using regular hexahedral elements. These elements are inherently limited in their ability to accurately conform to the curved boundaries of TPMS geometries. Consequently, when voxel-based meshing is applied to TPMS structures, a large number of elements is often required to achieve reliable numerical results, which significantly reduces computational efficiency.

To solve the above problems, this study introduces the minimum Jacobian (MJ) as a control parameter for voxel element generation in TPMS structures. The grid convergence index (GCI) based on Richardson extrapolation is introduced to evaluate the numerical asymptotic solution and mesh convergence behavior of TPMS models. The rest of the paper is arranged as follows. In Section 2, the two-parameter voxel modeling method is introduced, in which both the element size and the MJ are jointly controlled. The corresponding numerical simulation model is established based on this approach. In Section 3, the numerical convergence is analyzed using the relative error method and Richardson extrapolation, and the superiority of the two-parameter voxel method is verified. In Section 4, the applicability of the proposed method to the analysis of graded TPMS structures is further validated. Finally, the conclusions are drawn in Section 5.

## 2.Methodology

### 2.1. Design of TPMS structures

Various mathematical representations of the TPMS have been proposed [28], typically expressed more conveniently using an implicit function in the Cartesian coordinate system. The Gyroid surface can be represented as follows:

$$\varphi_G = \sin X \times \cos Y + \sin Y \times \cos Z + \sin Z \times \cos X = C \tag{1}$$

where $x = 2\pi a / L_x$, $y = 2\pi b / L_y$, $z = 2\pi c / L_z$, the parameters $a$, $b$ and $c$ control the dimensions of the entire structure in the x-, y-, and z-directions, respectively. The surface offset is determined by parameter $C$. The degree of surface offset determines the relative density of a TPMS structure. Relative density (RD) is defined as the ratio of the volume of the solid region to the volume of the whole cube occupied by the structure, which can also be expressed as [29]:

$$\mathrm{RD} = \frac{\rho}{\rho_s} = 1 - \varphi \tag{2}$$

where $\rho$ is the equivalent density of the TPMS structure, $\rho_s$ is the density of the base material, and $\varphi$ is the porosity of the structure.

TPMS structures can be generated by directly offsetting the corresponding TPMS surfaces. As shown in Fig. 1, the structure obtained by applying an offset to the Gyroid surface is called sheet-network Gyroid structure [30-31] or Double Gyroid (DG) structure [32-33]. The two domains of the cubic space divided by the sheet-network Gyroid are called solid-network Gyroid structure [31] or Gyroid structure. In this paper, the Gyroid strut-based structure is selected for the study and the STL file of the TPMS structure is generated by MATLAB code.

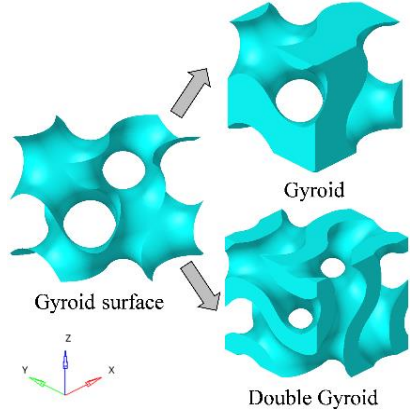

Fig. 1. The Gyroid surface are offset to generage Double Gyroid (DG) and Gyroid structures.

### 2.2. Two-parameter voxel model

#### 2.2.1. Isoparametric transformation of hexahedral elements

In the finite element method, each element is defined within its own local coordinate system, also known as the natural coordinate system. However, numerical computations and the interpretation of physical phenomena are conducted in the global coordinate system. To map physical quantities from the local system to the global system, a transformation matrix is required. The Jacobian matrix serves this purpose by describing the geometric transformation between the local and global coordinate systems.

Assuming that the local coordinates of the hexahedral element are denoted as $(\xi, \eta, \zeta)$, satisfying $\xi, \eta, \zeta \in [-1,1]$. The form function $N_i^{'}(\xi, \eta, \zeta)$ of the eight-node hexahedral element in local coordinates can be expressed as:

$$N_i^{'}(\xi, \eta, \zeta) = \frac{1}{8}(1 + \xi \xi_j)(1 \pm \eta \eta_j)(1 \pm \zeta \zeta_j) \tag{3}$$

where the values of $\xi_j$, $\eta_j$ and $\zeta_j$ are $\pm 1$. The global coordinates of any point within the element can be expressed as an interpolation of the nodal coordinates using the shape functions:

$$\begin{cases} x = \sum_{i=1}^{m} N_i^{'} x_i \\ y = \sum_{i=1}^{m} N_i^{'} y_i \\ z = \sum_{i=1}^{m} N_i^{'} z_i \end{cases} \tag{4}$$

where $m$ is the number of nodes for which the coordinate transformation is performed, $x_i$, $y_i$, and $z_i$ are the coordinate values of the nodes within the overall coordinates, and $N_i^{'}$ is the shape function expressed in local coordinates. It can be obtained by differentiation:

$$\begin{bmatrix} \dfrac{\partial N_i}{\partial \xi} \\[2mm] \dfrac{\partial N_i}{\partial \eta} \\[2mm] \dfrac{\partial N_i}{\partial \zeta} \end{bmatrix} = \begin{bmatrix} \dfrac{\partial x}{\partial \xi} & \dfrac{\partial y}{\partial \xi} & \dfrac{\partial z}{\partial \xi} \\[2mm] \dfrac{\partial x}{\partial \eta} & \dfrac{\partial y}{\partial \eta} & \dfrac{\partial z}{\partial \eta} \\[2mm] \dfrac{\partial x}{\partial \zeta} & \dfrac{\partial y}{\partial \zeta} & \dfrac{\partial z}{\partial \zeta} \end{bmatrix} \begin{bmatrix} \dfrac{\partial N_i}{\partial x} \\[2mm] \dfrac{\partial N_i}{\partial y} \\[2mm] \dfrac{\partial N_i}{\partial z} \end{bmatrix} = \mathbf{J} \begin{bmatrix} \dfrac{\partial N_i}{\partial x} \\[2mm] \dfrac{\partial N_i}{\partial y} \\[2mm] \dfrac{\partial N_i}{\partial z} \end{bmatrix} \tag{5}$$

where $\mathbf{J}$ is the Jacobian matrix. Accordingly, the relationship between the shape functions in the local and global coordinate systems can be expressed as:

$$\begin{bmatrix} \dfrac{\partial N_i}{\partial x} \\[2mm] \dfrac{\partial N_i}{\partial y} \\[2mm] \dfrac{\partial N_i}{\partial z} \end{bmatrix} = \mathbf{J}^{-1} \begin{bmatrix} \dfrac{\partial N_i}{\partial \xi} \\[2mm] \dfrac{\partial N_i}{\partial \eta} \\[2mm] \dfrac{\partial N_i}{\partial \zeta} \end{bmatrix} \tag{6}$$

where $\mathbf{J}^{-1}$ is the inverse matrix of $\mathbf{J}$ and $|\mathbf{J}|$ is the Jacobian determinant. The Jacobian can be interpreted as the ratio of a unit volume defined in the local (natural) coordinate system to its transformed volume in the global coordinate system. A relatively small Jacobian value typically indicates a high level of geometric distortion or element warping.

### 2.2.2. Two-parameter voxel modeling method

Fig. 2 shows the proposed two-parameter voxel method. By simultaneously reducing the element size and the MJ, the finite element model progressively approximates the underlying TPMS geometry, thereby enabling the numerical simulation to converge toward the reference solution. The two-parameter voxel model can be implemented by the commercial software Hypermesh. In this study, the implicit representation of the TPMS structure is first constructed in MATLAB, and the corresponding STL file is generated. The STL model is then imported into Hypermesh, where the Shrink Wrap module is employed to discretize the geometry into voxel elements, and the minimum Jacobian module is used to regulate the MJ.

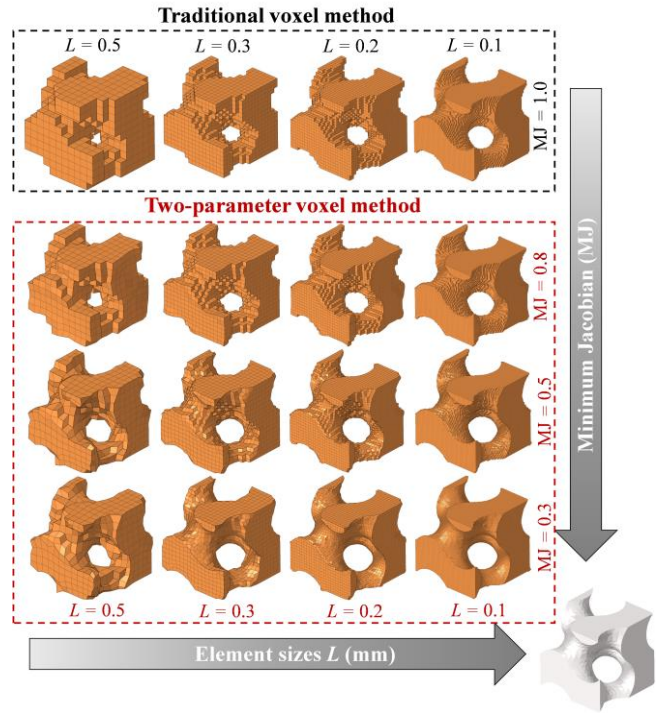

Fig. 2. Overview of the traditional voxel method and proposed two-parameter voxel method.

### 2.3. Numerical simulation model

Several studies have demonstrated that the TPMS structure exhibit significant size effects. When the lattice reaches at least four unit cells in each of the three orthogonal directions [16-17], the structure tends to exhibit stable and representative mechanical behavior. Fig. 3 shows the quasi-static compression simulation model of a 4×4×4 TPMS structure with a relative density of 0.45 and a unit cell size of $L_0 = 5$ mm. Two rigid walls are applied to simplify the loading system. The lower rigid plate is fully fixed, while the upper plate moves downward by 0.1 mm at a constant speed of 20 mm/s. The friction coefficient between the rigid plates and the TPMS structure is set to 0.1 [34]. The equivalent stresses of the Gyroid structure are calculated using the following equation:

$$\sigma_e = \frac{F}{S} \tag{7}$$

where $F$ is the rigid wall force and $S$ is the initial nominal cross-sectional area of the structure. Ti-6Al-4V (TC4) is used as the base material for the TPMS structure, and its mechanical properties are represented by the elasto-plastic linear hardening constitutive model, which exhibits bilinear stress-strain characteristics [35]. The material parameters used in the numerical simulations are listed in Table 1 and are derived from our previous work [36].

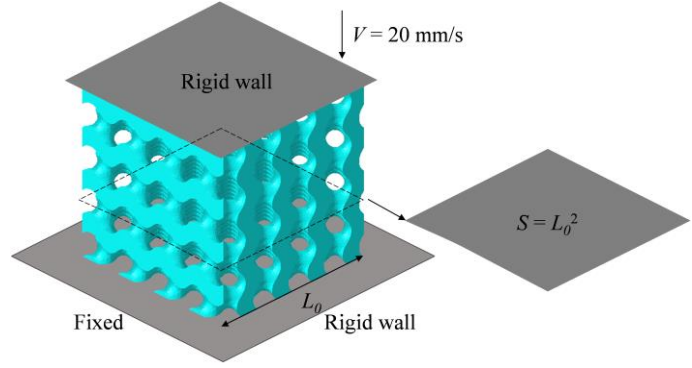

Fig. 3. Numerical simulation model of the Gyriod structure, where $S$ is the nominal cross-sectional area.

Table 1   Material parameters of Ti-6Al-4V.

| Parameter | Density $\rho_s$ (kg/m³) | Young's modulus $E_s$ (GPa) | Poisson's ratio $v$ | Yield stress $\sigma_y$ (MPa) | Tangent modulus $E_t$ (GPa) |
|---|---|---|---|---|---|
| Value | 4400 | 121 | 0.34 | 896 | 1.85 |

## 3. Mesh convergence analysis of uniform TPMS

### 3.1. The influence of minimum Jacobian on equivalent Young's modulus

Fig. 4 (a) and Fig. 4 (b) respectively show the discretization results of the Gyroid structure in the finite element model under constant element size and constant MJ conditions. The orange region represents the finite element model of the Gyroid structure, while the red region corresponds to its geometric model. As shown in Fig. 4, both decreasing the element size and MJ can improve the fidelity of the finite element mesh in approximating the geometric model of the Gyroid structure. In addition, when MJ=1, the finite element model extends beyond the geometric boundary regardless of the element size. In contrast, when MJ ≤ 0.3, the number of elements stabilizes, and the finite element mesh exhibits a significantly improved fit to the Gyroid geometry.

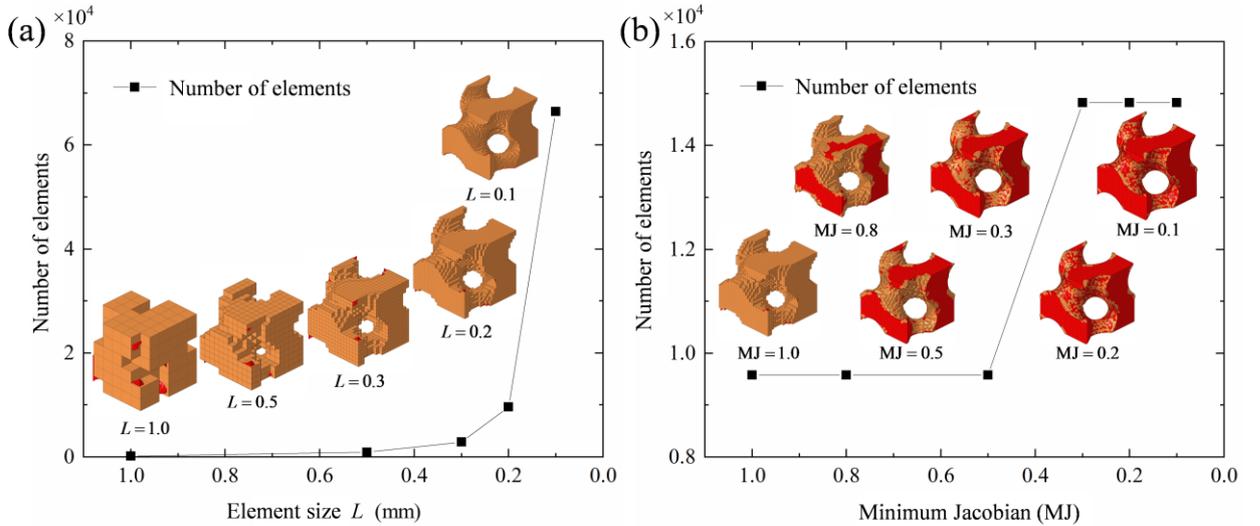

Fig. 4. Gyroid structures with a relative density of 0.45, where the red region represents the geometric model and orange region represents the finite element model. (a) Models with Jacobian = 1 and different element sizes. (b) Models with element sizes of 0.2 mm and different MJ.

Different finite element models of the Gyroid structure were used to investigate the effect of MJ on simulation results. Fig. 5 shows the effective Young's modulus for various MJ values at a fixed element size of 0.2 mm. The results indicate that the modulus stabilizes when MJ ≤ 0.3. Therefore, MJ=0.3 is adopted for mesh generation in this study.

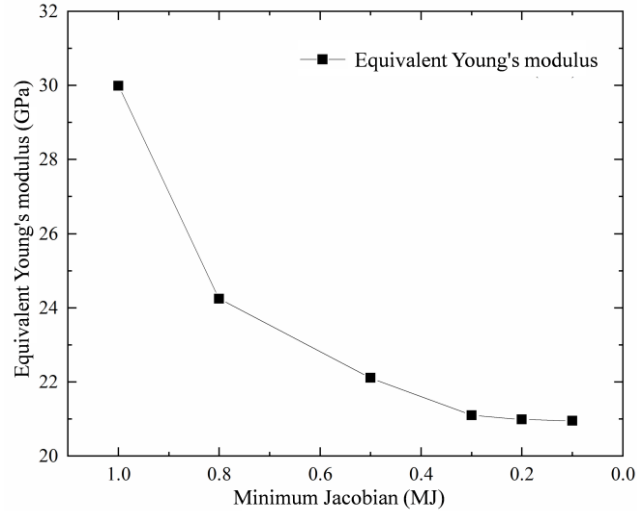

Fig. 5. Numerical simulation results of equivalent Young's modulus for models with different MJ.

### 3.2. Mesh convergence analysis

Fig. 6 shows the number of elements per unit cell and equivalent Young's modulus for both the voxel and two-parameter voxel models of the Gyroid structure under different element sizes. At a given element size, the two-parameter voxel model requires a greater number of elements to discretize the TPMS geometry.

Numerical results indicate that both modeling methods exhibit a decreasing trend in effective Young's modulus with mesh refinement. Within the element size range of 0.4 mm to 0.1 mm, the effective Young's modulus of the voxel model exhibits an approximately linear dependence on element size, whereas the modulus for the two-parameter model gradually stabilizes. Moreover, for the same element size, the voxel model yields a significantly higher effective Young's modulus than the two-parameter model. This is attributed to the fact that the voxel mesh extends significantly beyond the geometric boundary of the TPMS structure, thereby resulting in finite element models whose relative density exceeds the intended design value.

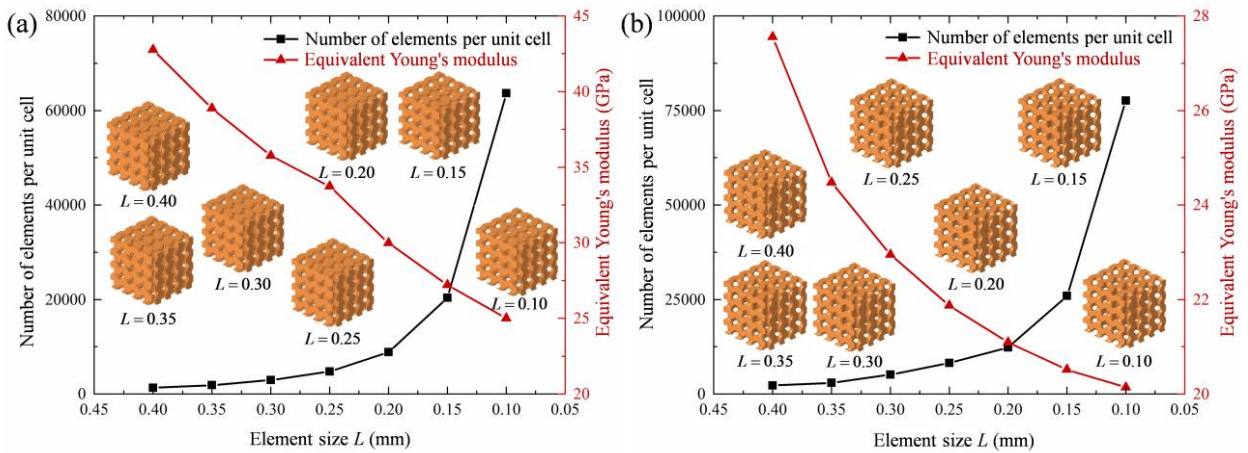

Fig. 6. Number of elements per unit cell and equivalent Young's modulus of Gyroid structures with different element sizes. (a) Voxel model. (b) Two-parameter voxel model (MJ=0.3).

### 3.2.1.Relative error method

The relative error method evaluates convergence by calculating the relative error between numerical results obtained from neighbouring element size. If the error between successive mesh refinements falls below the predefined value, the mesh is considered to have reached convergence. The relative error $\varepsilon$ can be expressed as:

$$\varepsilon = \frac{|f_2 - f_1|}{f_1} \times 100\% \tag{8}$$

where $f_1$ and $f_2$ are the numerical simulation results for finer and coarser meshes, respectively. Table 2 shows the equivalent Young's modulus and relative errors of the two modeling methods. Within the mesh size range of 0.4 mm to 0.1 mm, the voxel model shows no significant reduction in relative error. In contrast, the proposed two-parameter voxel model exhibits a progressively decreasing relative error with mesh refinement. When the element size reaches 0.1 mm, the relative error in the effective Young's modulus is reduced to just 1.87%.

Table 2   Number of single-cell elements, equivalent Young's modulus and relative error for two TPMS models.

| | Element size (mm) | 0.4 | 0.35 | 0.3 | 0.25 | 0.2 | 0.15 | 0.1 |
|---|---|---|---|---|---|---|---|---|
| MJ=1 | Number of elements per unit cell | 1328.8 | 1883.2 | 2947.0 | 4776.0 | 8836.7 | 20392.2 | 63632.8 |
| | Equivalent Young's modulus (GPa) | 42.756 | 38.901 | 35.756 | 33.72 | 29.986 | 27.21 | 25.004 |
| | Relative error $\varepsilon$ (%) | / | 9.91 | 8.80 | 6.04 | 12.45 | 10.20 | 8.82 |
| MJ=0.3 | Number of elements per unit cell | 2208.8 | 3022.8 | 5116.08 | 8208.0 | 12348.9 | 25984.0 | 77662.8 |
| | Equivalent Young's modulus (GPa) | 27.553 | 24.474 | 22.95 | 21.874 | 21.093 | 20.514 | 20.137 |
| | Relative error $\varepsilon$ (%) | / | 12.58 | 6.64 | 4.92 | 3.70 | 2.82 | 1.87 |

### 3.2.2.Richardson extrapolation method

The GCI is another indicator to assess the degree of convergence of the numerical simulation results with mesh refinement. The GCI is calculated based on the Richardson extrapolation method, which was originally developed for applications in computational fluid dynamics. In recent years, this approach has been increasingly adopted in the context of finite element analysis for solid and structural mechanics. In this method, the discretization error of the finite element solution $\tilde{E}$ is assumed as [37]:

$$\tilde{E} = f_h - f_{exact} = Ch^p + H.O.T. \tag{9}$$

where $\tilde{E}$ is the difference between the result of the current mesh density $f_h$ and the exact solution $f_{exact}$, $C$ is a constant, and $H.O.T.$ denotes the higher order terms. $p$ is the order of convergence, which can be expressed as:

$$p = \frac{\ln\left(\dfrac{f_3 - f_2}{f_2 - f_1}\right)}{\ln(r)} \tag{10}$$

where $f_1$, $f_2$ and $f_3$ are the numerical simulation results at three different element sizes $h_1$, $h_2$ and $h_3$, respectively, and $r$ is the grid refinement ratio.

Richardson extrapolation is used to evaluate the higher order estimates of the numerical simulation results. The numerical simulation result $f$ for element size $h$ can be expressed using Taylor's theorem as:

$$f = f_{h=0} + g_1 h + g_2 h^2 + g_3 h^3 + ... \tag{11}$$

where $f_{h=0}$ is the asymptotic solution for $h$ approaching 0. The unknown parameters $g_1$, $g_2$ and $g_3$ are independent of the element sizes $h$. For two element sizes $h_1$ and $h_2$, ignoring third and higher order terms, the asymptotic solution can be expressed as:

$$f_{h=0} \cong f_1 + \frac{f_1 - f_2}{r^p - 1} \tag{12}$$

The GCI is defined by the following equation [37]:

$$\text{GCI} = \frac{F_s |\varepsilon|}{r^p - 1} \times 100\% \tag{13}$$

where $F_s$ is the factor of safety, $F_s = 3$ is used when two sets of element sizes are used and $F_s = 1.25$ for three and more sets of element sizes. In order to ensure the accuracy of the GCI calculations, all calculations must be performed within the range of asymptotic convergence. This can be checked by comparing the GCI values to satisfy the following equation:

$$r^p \text{GCI}_{12} = \text{GCI}_{23} \tag{14}$$

The effective Young's modulus results obtained from mesh sizes of 0.1 mm, 0.2 mm, and 0.4 mm are used to calculate the GCI, with the corresponding values denoted as $f_1$, $f_2$, and $f_3$, respectively. The grid refinement ratio $r = \dfrac{f_2}{f_1} = \dfrac{f_3}{f_2} = 2$. For each mesh group, the order of convergence $p$, the asymptotic solution $f_{h=0}$, the $\mathrm{GCI}_{12}$ between 0.1 mm, 0.2 mm elements and $\mathrm{GCI}_{23}$ between 0.2 mm, 0.4 mm elements are calculated. Finally, the ratio

$R_a = \dfrac{\mathrm{GCI}_{23}}{r^p \mathrm{GCI}_{12}}$, defined by (14), is evaluated. When $R_a$ is approximately equal to 1, it indicates that the numerical solution is within the asymptotic range of convergence.

Table 3  Asymptotic solutions and GCI for two TPMS structural modeling methods.

| Modeling method | Order of convergence $p$ | Asymptotic solution $f_{h=0}$ (GPa) | $\mathrm{GCI}_{12}$ (%) | $\mathrm{GCI}_{23}$ (%) | $R_a$ |
|---|---|---|---|---|---|
| Voxel model | 1.358 | 21.817 | 15.932 | 40.838 | 0.99999 |
| Two-parameter voxel model | 2.756 | 19.97 | 1.031 | 6.965 | 1.00005 |

Table 3 shows the GCI results for both modeling methods. The values of $R_a$ for both models are close to 1, indicating that the numerical simulation results are in the asymptotic range of convergence. At the same element sizes, the two-parameter voxel model exhibits lower relative error, higher order of convergence and lower GCIs compared to the voxel model. These results indicate that the proposed method offers a significant advantage in terms of mesh convergence performance.

### 3.3. Analysis of solution accuracy

The Gibson–Ashby model is widely used to describe the relationship between the relative density and mechanical properties of porous structures. The elastic modulus model based on the Gibson–Ashby model can be expressed as [38]:

$$\frac{E}{E_s} = C_1 (\frac{\rho}{\rho_s})^m \tag{15}$$

where $E$ and $\rho$ are the elastic modulus and density of TPMS structures, $E_s$ and $\rho_s$ are the elastic modulus and density of its base material, respectively. $E/E_s$ is the relative modulus. $C_1$ and $m$ are both fitting parameters.

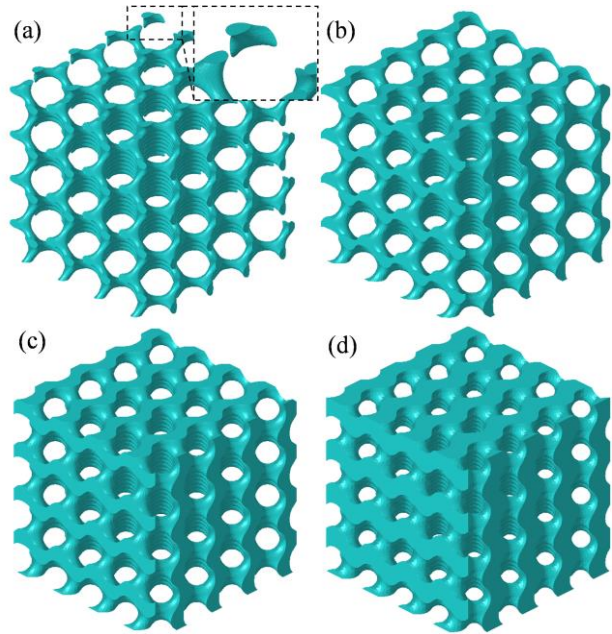

Fig. 7. Geometric models of Gyroid structures with relative density of (a) 0.1, (b) 0.2, (c) 0.3 and (d) 0.45, with deletion of the discontinuous part of the Gyroid with RD=0.1.

Fig. 7 shows the geometrical model of the Gyroid structure with relative density of 0.1, 0.2, 0.3, and 0.45, respectively. For the structure with a relative density of 0.1, discontinuous regions are observed. In this study, all these models are executed on a computer with an i7-12700F@2.10 GHz CPU and 32 GB RAM. Considering the computational time required for both preprocessing and solving, the 0.15 mm voxel model and the 0.2 mm two-parameter voxel model are selected for numerical simulation in this section. At a relative density of 0.45, the corresponding solving times are 121 s and 58 s, respectively. Fitting Eq. (15), the relationship between relative modulus and relative density of Gyroid structure can be expressed as:

$$\frac{E}{E_s} = 1.11(\frac{\rho}{\rho_s})^{1.96} \tag{16}$$

$$\frac{E}{E_s} = 1.06(\frac{\rho}{\rho_s})^{2.24} \tag{17}$$

Eqs. (16) and (17) show the results of the voxel models with $L$=0.15 mm and the two-parameter voxel models with $L$=0.2 mm, respectively. Both methods fit the Gibson-Ashby model well, with coefficients of determination $R^2 > 0.99$.

Fig. 8 presents a comparison between the numerical results obtained in this study and previously reported experimental and numerical results [39-41]. The two-parameter voxel model with an element size of 0.2 mm shows good consistency with the numerical results of Peng et al. [39] (4 mm lattices with 0.15 mm tetrahedral elements), as well as with the results reported by Yang et al. [40] (unspecified modeling method). Moreover, the results from the two-parameter voxel model are closer to the experimental data compared to those from the conventional voxel model. These findings demonstrate that, compared to the traditional voxel method, the proposed method offers advantages in both computational efficiency and accuracy.

Fig. 8 also shows that the experimentally measured Young's modulus is generally lower than the numerical results. This discrepancy can be attributed to three main factors:

(1) Most TPMS specimens are currently fabricated using additive manufacturing. For thin-walled or small-scale TPMS structures, the additive manufacturing process inevitably introduces initial defects in the structure [42], which in turn reduce the actual Young's modulus.

(2) Variations in heat treatment and manufacturing processes may lead to differences in the material's microstructure, which can subsequently affect the measured Young's modulus.

(3) Post-processing methods can also contribute to discrepancies between experimental results and numerical simulations. According to existing studies, sandblasting can remove surface impurities and defects, thereby improving the surface quality of TPMS structures and enhancing their mechanical performance [43].

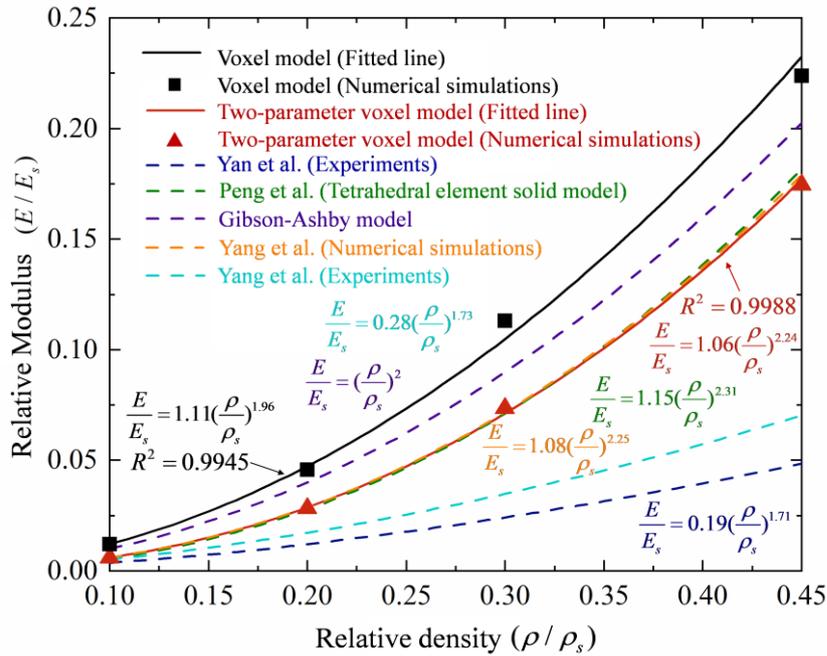

Fig. 8. Gibson-Ashby model of the relative modulus of Gyroid structure, compared with the experimental and numerical simulation results of Peng et al [39], Yang et al [40], and Yan et al [41].

## 4. Mesh convergence analysis of graded TPMS

### 4.1. Graded TPMS model

To further enhance the mechanical performance of TPMS structures, the graded structure strategy has been recently introduced to TPMS structures. By continuously or stepwise varying sheet thickness or cell size, functionally graded TPMS structures can be tailored to specific application needs [44-46]. In this section, the applicability of the proposed two-parameter voxel method to graded TPMS structures is evaluated using a graded Gyroid structure, as illustrated in

Fig. 9. The structure exhibits a gradual transition in relative density from 0.35 at the top to 0.55 at the bottom, resulting in an overall relative density of 0.45. The same rigid wall model shown in Fig. 3 is used to apply compression to the graded TPMS.

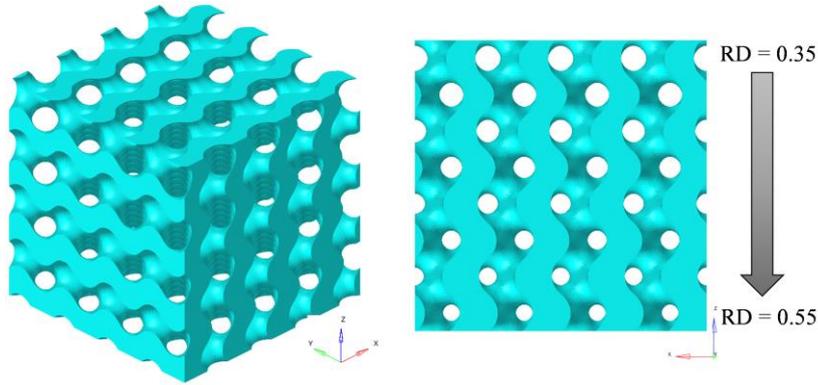

Fig. 9. The geometric model of the graded Gyroid structure with relative density showing a uniform transition from 0.35 to 0.55 from top to bottom.

### 4.2. Mesh convergence analysis

Fig. 10 shows the effective Young's modulus of the graded TPMS structure under different element sizes for both the voxel model and the two-parameter voxel model. The asymptotic solutions obtained from both modeling methods are slightly lower than those of the uniform structure with the same overall relative density. This discrepancy may be attributed to the fact that the low-density regions in the graded Gyroid structure are more prone to deformation. Table 4 shows the relative errors of the two voxel modeling methods, and Table 5 shows the results of the calculation of the order of convergence, the GCI and the asymptotic solutions.

The numerical results show that, similar to the case of uniform TPMS, the voxel graded TPMS models with element sizes ranging from 0.4 mm to 0.1 mm also fail to produce accurate results. However, the two-parameter voxel model demonstrates a higher order of convergence and a lower GCI. When the element size is reduced to 0.1 mm, the numerical results closely match the asymptotic solution. These findings show that the two-parameter voxel method proposed in this paper is also applicable to graded TPMS structures.

In this section, the $\text{GCI}_{12}$ and $\text{GCI}_{23}$ of the graded TPMS are 0.029% and 1.815%, respectively, both lower than the 1.031% and 6.965% of the uniform TPMS. When the element size is smaller than 0.25 mm, the relative errors $\varepsilon$ remain below the corresponding values observed for the homogeneous TPMS structure with MJ=0.3. These results suggest that the two-parameter voxel method may offer superior mesh convergence performance when applied to graded TPMS structures compared to their uniform counterparts.

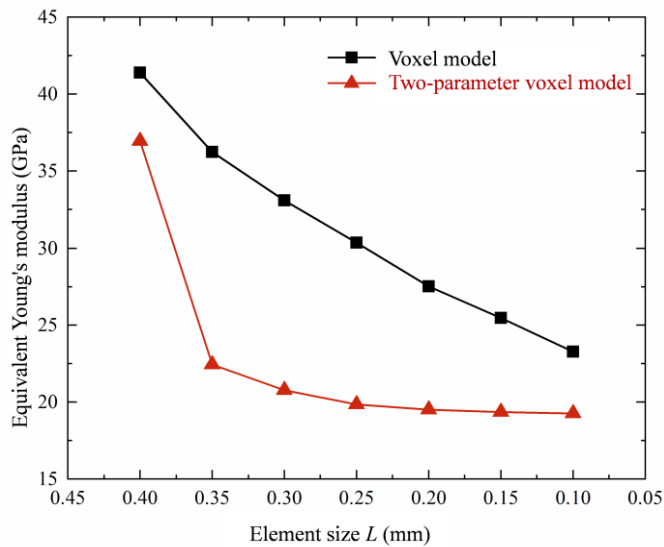

Fig. 10. Equivalent Young's modulus of graded Gyroid structures with different element sizes.

Table 4   Number of elements, equivalent Young's modulus and relative error for two TPMS models.

| | Element sizes (mm) | 0.4 | 0.35 | 0.3 | 0.25 | 0.2 | 0.15 | 0.1 |
|---|---|---|---|---|---|---|---|---|
| MJ = 1 | Number of elements | 83930 | 119315 | 177354 | 300174 | 558953 | 1250272 | 4026664 |
| | Equivalent Young's modulus (GPa) | 41.40 | 36.23 | 33.10 | 30.35 | 27.53 | 25.45 | 23.25 |
| | Relative error $\varepsilon$ (%) | / | 12.48 | 8.63 | 8.31 | 9.30 | 7.56 | 8.64 |
| MJ = 0.3 | Number of elements | 139034 | 191909 | 274946 | 444322 | 784157 | 1646896 | 4919576 |
| | Equivalent Young's modulus (GPa) | 36.94 | 22.43 | 20.77 | 19.85 | 19.52 | 19.36 | 19.24 |
| | Relative error $\varepsilon$ (%) | / | 39.28 | 7.40 | 4.43 | 1.66 | 0.82 | 0.62 |

Table 5   Asymptotic solutions and GCI for two TPMS structural modeling methods.

| Modeling method | Order of convergence $p$ | Asymptotic solution (GPa) | $GCI_{12}$ (%) | $GCI_{23}$ (%) | $R_a$ |
|---|---|---|---|---|---|
| Voxel model | 1.696 | 21.34 | 10.27 | 33.28 | 1.0002 |
| Two-parameter voxel model | 5.986 | 19.24 | 0.029 | 1.815 | 0.987 |

## 5.Conclusions

This study proposes a finite element modeling method for TPMS structures based on voxelization, incorporating joint regulation of element size and MJ. The numerical convergence behavior of both voxel models and the proposed two-parameter voxel models is systematically evaluated using relative error method and Richardson extrapolation. Finally, the feasibility of applying the two-parameter voxel method to the modeling of graded TPMS structures is verified. The main conclusions are as follows:

(1) The CGI method based on Richardson extrapolation is introduced for the verification of numerical models of TPMS structures. This approach provides a systematic and objective framework for evaluating discretization errors and conducting mesh convergence analysis in voxel TPMS modeling.

(2) A two-parameter voxel method is proposed for finite element modeling of TPMS structures, in which element size and the MJ are jointly controlled to optimize geometric approximation accuracy. This method retains the same level of convenience as the conventional voxel method. When the MJ value is reduced below 0.3, the numerical results tend to stabilize. Numerical results indicate that, compared to the conventional voxel method, the proposed two-parameter voxel method exhibits superior mesh convergence, higher computational efficiency, and improved solution accuracy.

(3) The two-parameter voxel method also demonstrates significant advantages in the modeling of graded TPMS structures. In the present case study, it exhibits superior convergence performance compared to its application in uniform TPMS structures.

Despite the demonstrated advantages of the proposed method, certain limitations remain when addressing nonlinear problems involving impact or large deformations, which typically require explicit time integration. Specifically, when the MJ=0.3, some elements exhibit excessively small characteristic lengths, leading to a substantial reduction in the stable time step and, consequently, a significant increase in computational time. As a result, models with MJ=0.3 are currently unsuitable for explicit analyses. Future research will focus on optimizing the MJ to achieve a better balance between numerical accuracy and computational efficiency in explicit simulations of TPMS structures.

**Declaration of competing interest**

The authors declare that they have no known competing financial interests or personal relationships that could have appeared to influence the work reported in this paper.

**Data availability**

Data will be made available on request.

**Acknowledgments**

The authors acknowledged the financial supports from National Key Research and Development Program (No. 2021YFB3302200).

# References

[1] H. Yin, W. Zhang, L. Zhu, F. Meng, J. Liu, G. Wen, Review on lattice structures for energy absorption properties, Compos. Struct. 304 (2023) 116397.

[2] N. Qiu, Y. Wan, Y. Shen, J. Fang, Experimental and numerical studies on mechanical properties of TPMS structures, Int. J. Mech. Sci. 261 (2024) 108657.

[3] C. Peng, K. Fox, M. Qian, H. Nguyen-Xuan, P. Tran, 3D printed sandwich beams with bioinspired cores: Mechanical performance and modelling, Thin-Walled Struct. 161 (2021) 107471

[4] O. Al-Ketan, R. Rowshan, R.K.A. Al-Rub, Topology–mechanical property relationship of 3D printed strut, skeletal, and sheet based periodic metallic cellular materials, Addit. Manuf. 19 (2018) 167–183.

[5] H. Yin, Z. Liu, J. Dai, G. Wen, C. Zhang, Crushing behavior and optimization of sheet-based 3D periodic cellular structures, Compos. Part B Eng. 182 (2020) 107565.

[6] S. Li, M. Hu, L. Xiao, W. Song, Compressive properties and collapse behavior of additively-manufactured layered-hybrid lattice structures under static and dynamic loadings, Thin-Walled Struct. 157 (2020) 107153.

[7] L. Xiao, W. Song, C. Wang, H. Tang, N. Liu, J. Wang, Yield behavior of open-cell rhombic dodecahedron Ti–6Al–4V lattice at elevated temperatures, Int. J. Mech. Sci. 115 (2016) 310–317.

[8] J.B. Berger, H.N. Wadley, R.M. McMeeking, Mechanical metamaterials at the theoretical limit of isotropic elastic stiffness, Nature 543 (2017) 533–537.

[9] N. Novak, S. Tanaka, K. Hokamoto, A. Mauko, Y.E. Yilmaz, O. Al-Ketan, Z. Ren, High strain rate mechanical behaviour of uniform and hybrid metallic TPMS cellular structures, Thin-Walled Struct. 191 (2023) 111109.

[10] L. Wan, D. Hu, M. Wan, Z. Yang, H. Zhang, B. Pi, Lateral crushing behavior of tubular lattice structures with triply periodic minimal surface architectures, Thin-Walled Struct. 189 (2023) 110905.

[11] Y. Zhang, Y. Chen, J. Li, J. Wu, L. Qian, Y. Tan, G. Zeng, Protective performance of hybrid triply periodic minimal surface lattice structure, Thin-Walled Struct. 194 (2024) 111288.

[12] R. Santiago, H. Ramos, S. AlMahri, O. Banabila, H. Alabdouli, D.W. Lee, A. Aziz, N. Rajput, M. Alves, Z. Guan, Modelling and optimisation of TPMS-based lattices subjected to high strain-rate impact loadings, Int. J. Impact Eng. 177 (2023) 104592.

[13] Z. Luo, Q. Tang, Q. Feng, S. Ma, J. Song, R. Setchi, F. Guo, Y. Zhang, Finite element analysis of the mechanical properties of sheet-and skeleton-gyroid Ti6Al4V structures produced by laser powder bed fusion, Thin-Walled Struct. 192 (2023) 111098.

[14] X. Li, L. Xiao, W. Song, Compressive behavior of selective laser melting printed Gyroid structures under dynamic loading, Addit. Manuf. 46 (2021) 102054.

[15] S. Li, E. Xu, X. Zhan, W. Zheng, L. Li, Stress-driven nonlocal homogenization method for cellular structures, Aerosp. Sci. Technol. 155 (2024) 109632.

[16] I. Maskery, A.O. Aremu, L. Parry, R.D. Wildman, C.J. Tuck, I.A. Ashcroft, Effective design and simulation of surface-based lattice structures featuring volume fraction and cell type grading, Mater. Des. 155 (2018) 220–232.

[17] I. Maskery, L. Sturm, A.O. Aremu, A. Panesar, C.B. Williams, C.J. Tuck, R.D. Wildman, I.A. Ashcroft, R.J.M. Hague, Insights into the mechanical properties of several triply periodic minimal surface lattice structures made by polymer additive manufacturing, Polym. 152 (2018) 62–71.

[18] H. Montazerian, E. Davoodi, M. Asadi-Eydivand, J. Kadkhodapour, M. Solati-Hashjin, Porous scaffold internal architecture design based on minimal surfaces: a compromise between permeability and elastic properties, Mater. Des. 126 (2017) 98–114.

[19] A.O. Aremu, I. Maskery, C. Tuck, I.A. Ashcroft, R.D. Wildman, R.J.M. Hague, A comparative finite element study of cubic unit cells for selective laser melting, in: Proc. Int. Solid Freeform Fabr. Symp., Univ. Texas at Austin, 2014.

[20] Y. Wan, N. Qiu, M. Xiao, Y. Xu, J. Fang, Energy dissipation of 3D-printed TPMS lattices under cyclic loading, Int. J. Mech. Sci. 294 (2025) 110245.

[21] M. Afshar, A.P. Anaraki, H. Montazerian, J. Kadkhodapour, Additive manufacturing and mechanical characterization of graded porosity scaffolds designed based on triply periodic minimal surface architectures, J. Mech. Behav. Biomed. Mater. 62 (2016) 481–494.

[22] J. Kadkhodapour, H. Montazerian, A.C. Darabi, A. Zargarian, S. Schmauder, The relationships between deformation mechanisms and mechanical properties of additively manufactured porous biomaterials, J. Mech. Behav. Biomed. Mater. 70 (2017) 28–42.

[23] W. Jiang, W. Liao, T. Liu, X. Shi, C. Wang, J. Qi, Y. Chen, Z. Wang, C. Zhang, A voxel-based method of multiscale mechanical property optimization for the design of graded TPMS structures, Mater. Des. 204 (2021) 109655.

[24] Wang Y, Li X, Yan Z, Du Y, Bai J, Liu B, Rabczuk T, Liu Y. HomoGenius: a Foundation Model of Homogenization for Rapid Prediction of Effective Mechanical Properties using Neural Operators. arXiv preprint arXiv:2404.07943. 2024.

[25] U. Simsek, A. Akbulut, C.E. Gayir, C. Basaran, P. Sendur, Modal characterization of additively manufactured TPMS structures: comparison between different modeling methods, Int. J. Adv. Manuf. Technol. 115 (2021) 657–674.

[26] L.E. Schwer, Guide for Verification and Validation in Computational Solid Mechanics, American Society of Mechanical Engineers Virginia, (2006).

[27] A.O. Aremu, J.P.J. Brennan-Craddock, A. Panesar, I.A. Ashcroft, R.J.M. Hague, R.D. Wildman, C. Tuck, A voxel-based method of constructing and skinning conformal and functionally graded lattice structures suitable for additive manufacturing, Addit. Manuf. 13 (2017) 1–13.

[28] M. Sarkari, S. Rahmati, M. Nikkhoo, Computer-aided porous scaffold design for tissue engineering using triply periodic minimal surfaces, Int. J. Precis. Eng. Manuf. 12 (1) (2011) 61–71.

[29] Y. Sun, Q. Li, Dynamic compressive behaviour of cellular materials: A review of phenomenon, mechanism and modelling, Int. J. Impact Eng. 112 (2018) 74–115.

[30] J. Zhang, S. Xie, T. Li, Z. Liu, S. Zheng, H. Zhou, A study of multi-stage energy absorption characteristics of hybrid sheet TPMS lattices, Thin-Walled Struct. 190 (2023) 110989.

[31] T.-D. Hoang, T.H. Ngo, K.Q. Tran, S. Li, H. Nguyen-Xuan, A stochastic multiscale asymptotic homogenization approach to 3D printed biodegradable resin TPMS bio-inspired structures, Thin-Walled Struct. 212 (2025) 113100.

[32] I. Maskery, N.T. Aboulkhair, A.O. Aremu, C.J. Tuck, I.A. Ashcroft, Compressive failure modes and energy absorption in additively manufactured double Gyroid lattices, Addit. Manuf. 16 (2017) 24–29.

[33] S.B.G.B. Sébastien, M. Werner, M. Hannula, S. Sharifi, G.P.R.L. Lajoinie, D. Eglin, J. Hyttinen, A.A. Poot, D.W. Grijpma, Surface curvature in triply-periodic minimal surface architectures as a distinct design parameter in preparing advanced tissue engineering scaffolds, Biofabrication 9 (2) (2017) 025001.

[34] S. Li, M. Hu, L. Xiao, W. Song, Compressive properties and collapse behavior of additively-manufactured layered-hybrid lattice structures under static and dynamic loadings, Thin-Walled Struct. 157 (2020) 107153.

[35] LS-Dyna Keyword User's Manual, Livermore, n.a.

[36] S. Wang, C. Jiang, C. Wang, B. Zhang, H. Luo, W. Feng, Bird strike resistance analysis for engine fan blade filled with triply periodic minimal surface, Aerosp. Sci. Technol. 161 (2025) 110109.

[37] O. Aycan, A. Topuz, L. Kadem, Evaluating uncertainties in CFD simulations of patient-specific aorta models using Grid Convergence Index method, Mech. Res. Commun. 133 (2023) 104188.

[38] L.J. Gibson, Mechanical behavior of metallic foams, Annu. Rev. Mater. Sci. 30 (2000) 191–227.

[39] X. Peng, Q. Huang, Y. Zhang, X. Zhang, T. Shen, H. Shu, Z. Jin, Elastic response of anisotropic Gyroid cellular structures under compression: Parametric analysis, Mater. Des. 205 (2021) 109706.

[40] L. Yang, C. Yan, C. Han, P. Chen, S. Yang, Y. Shi, Mechanical response of a triply periodic minimal surface cellular structures manufactured by selective laser melting, Int. J. Mech. Sci. 148 (2018) 149–157.

[41] C. Yan, L. Hao, A. Hussein, P. Young, Ti–6Al–4V triply periodic minimal surface structures for bone implants fabricated via selective laser melting, J. Mech. Behav. Biomed. Mater. 51 (2015) 61–73.

[42] Z. Zou, M. Simonelli, J. Katrib, G. Dimitrakis, R. Hague, Microstructure and tensile properties of additive manufactured Ti–6Al–4V with refined prior-β grain structure obtained by rapid heat treatment, Mater. Sci. Eng. A 814 (2021) 141271.

[43] L. Yang, C. Yan, W. Cao, Z. Liu, B. Song, S. Wen, C. Zhang, Y. Shi, S. Yang, Compression–compression fatigue behaviour of Gyroid-type triply periodic minimal surface porous structures fabricated by selective laser melting, Acta Mater. 181 (2019) 49–66.

[44] J. Zhang, X. Chen, Y. Sun, J. Yang, R. Chen, Y. Xiong, W. Hou, L. Bai, Design of a biomimetic graded TPMS scaffold with quantitatively adjustable pore size, Mater. Des. 218 (2022) 110665.

[45] H. Nguyen-Xuan, K.Q. Tran, C.H. Thai, J. Lee, Modelling of functionally graded triply periodic minimal surface (FG-TPMS) plates, Compos. Struct. 315 (2023) 116981.

[46] N. Novak, M. Borovinšek, O. Al-Ketan, Z. Ren, M. Vesenjak, Impact and blast resistance of uniform and graded sandwich panels with TPMS cellular structures, Compos. Struct. 300 (2022) 116174.